\documentstyle[aasms4,12pt,incite,psfig]{article}
\begin{document}

\newcommand{\ltsim}{\lower.5ex\hbox{$\; \buildrel < \over \sim \;$}}
\newcommand{\gtsim}{\lower.5ex\hbox{$\; \buildrel > \over \sim \;$}}
\newcommand{\order}[1]{\mbox{$\cal{O}$ ({#1})}}
\newcommand{\Pdot}{\mbox{$\dot {\rm P}$}}
\newcommand{\Pbdot}{\mbox{$\dot {\rm P}_{\rm b}$}}
\newcommand{\etal}{\mbox{\it et~al.}}
\newcommand{\suph}{\mbox{$^{\rm h}$}}
\newcommand{\supm}{\mbox{$^{\rm m}$}}
\newcommand{\sups}{\mbox{$^{\rm s}$}}
\def\arcmin{\hbox{$^\prime$}}
\def\arcsec{\hbox{$^{\prime\prime}$}}
\newcommand{\TEMPO}{{\sc tempo}}
%
%
\newdimen\digitwidth
{\catcode`?=\active
        \gdef\setmathspace   {
            \setbox0=\hbox{\rm0}
            \digitwidth=\wd0
            \catcode`?=\active
            \def?{\kern\digitwidth}
        }
}

\newcommand{\units}[1]{\mbox{$\rm\,#1$}}
\newcommand{\kpc}       {\mbox{\rm\,kpc}}
\newcommand{\Hz}        {\mbox{\rm\,Hz}}
\newcommand{\kHz}       {\mbox{\rm\,kHz}}
\newcommand{\MHz}       {\mbox{\rm\,MHz}}
\newcommand{\GHz}       {\mbox{\rm\,GHz}}
\newcommand{\Jy}        {\mbox{\rm\,Jy}}
\newcommand{\mJy}       {\mbox{\rm\,mJy}}
\newcommand{\uJy}       {\mbox{$\,\mu\rm{Jy}$}}
\newcommand{\K}         {\mbox{\rm\,K}}
\newcommand{\KperJy}    {\mbox{$\rm\,K\,Jy^{-1}$}}
\newcommand{\dm}        {\mbox{$\rm\,pc\,cm^{-3}$}}
\newcommand{\kms}       {\mbox{$\rm\,km\,s^{-1}$}}
\newcommand{\yr}        {\mbox{\rm\,y}}
\newcommand{\s}         {\mbox{\rm\,s}}
\newcommand{\ms}        {\mbox{\rm\,ms}}
\newcommand{\us}        {\mbox{$\,\mu\rm{s}$}}
\newcommand{\degyr}     {\mbox{$\,^\circ\,{\rm y}^{-1}$}}
\newcommand{\degree}    {\mbox{$^\circ$}}
\newcommand{\gauss}     {\mbox{\rm\,G}}
\newcommand{\accel}     {\mbox{$\rm\,m\,s^{-2}$}}
\newcommand{\Msun}      {\mbox{$\,M_{\mathord\odot}$}}
\newcommand{\Lsun}      {\mbox{$\,L_{\mathord\odot}$}}
\newcommand{\Rsun}      {\mbox{$\,R_{\mathord\odot}$}}

\newcommand{\lsi}{\mbox{LSI\,+61\degree\,303}}
\newcommand{\eg}{\mbox{\it e.g.~}}
\newcommand{\gt}{\mbox{GT~0236+610}}

\newcommand{\mt}{ } 

\title{Long-Term Flux Monitoring of \lsi\ at 2.25 and 8.3 GHz}
\author{P. S. Ray\altaffilmark{1}, R. S. Foster\altaffilmark{2}, 
E. B. Waltman\altaffilmark{3}}
\affil{ Code 7210 \\
	Remote Sensing Division \\
	Naval Research Laboratory \\
	Washington, DC~~20375}
\author{M. Tavani\altaffilmark{4}}
\affil{Columbia Astrophysics Laboratory\\
	Columbia University\\
	New York, NY~~10027}
\author{F. D. Ghigo\altaffilmark{5}}
\affil{National Radio Astronomy Observatory \\
	P.O. Box 2 \\
	Green Bank, WV~~24944}
\altaffiltext{1}{E-mail: paulr@rira.nrl.navy.mil, NRC Research Associate}
\altaffiltext{2}{E-mail: foster@rira.nrl.navy.mil}
\altaffiltext{3}{E-mail: waltmane@rira.nrl.navy.mil}
\altaffiltext{4}{E-mail: tavani@carmen.phys.columbia.edu}
\altaffiltext{5}{E-mail: fghigo@nrao.edu}
\begin{abstract}

\lsi\ is an exotic binary system consisting of a $\sim 10 \Msun$ B
star and a compact object which is probably a neutron star.  The
system {is associated with the interesting radio source \gt\ that}
exhibits bright radio outbursts {with a period of 26.5 days.}  We
report {the results of continuous} daily radio interferometric
observations of \gt\ at 2.25 and 8.3 GHz from 1994 January to 1996
February.  The observations cover 25 complete (and 3 partial) {cycles}
with multiple observations each day.  We detect {substantial
cycle-to-cycle variability of the radio emission characterized by a
rapid onset of the radio flares followed by a more gradual decrease of
the emission.  We detect a systematic change of the radio spectral
index $\alpha$ (defined as $S_{\nu} \propto \nu^{\alpha}$) which
typically becomes larger than zero at the onset of the radio
outbursts.}  This behavior is suggestive of expansion of material
initially optically thick to radio frequencies, indicating either that
synchrotron or inverse Compton cooling are important or that the
free-free optical depth to the source is rapidly changing.  After two
years of observations, we see only weak evidence for the proposed
4-year periodic modulation in the peak flux of the outbursts.  We
observe a secular trend in the outburst phases according the the best
published ephemeris.  This trend indicates either orbital period
evolution, or a drift in outburst orbital phase in response to some
other change in the system.
\end{abstract}


\section{Introduction}

The exotic binary system \lsi\ (= V615 Cas = GT\,0236+610) is one of a
remarkable class of X-ray and radio emitting binaries which includes
the well-known sources Cir X-1, Cyg X-3 and SS433 (\eg \cite{hj88}).
\lsi\ is a $\sim 10 \Msun$ B0 star at a distance of $\sim 2$\,kpc
(\cite{fh91}) whose spectrum shows evidence for rapid rotation and a
high-velocity equatorial wind (\cite{hc81}).  The B star is in a
binary system with a compact companion, most likely a neutron star.
This source is particularly {interesting} among high-mass binaries
because of its strongly variable emission at wavelengths from radio to
{X-rays and probably} $\gamma$-rays.  Although this source has been
the target of a great many observations, the energy source and
emission mechanisms responsible for its peculiar behavior remain a
mystery.

\gt\ is a highly variable radio source which exhibits periodic radio
outbursts with a period of $26.496 \pm 0.008$\,d (\cite{tg84}).
However, these outbursts are not stable in phase.  Outburst maxima
have been seen from phase 0.45 to 0.95, but bright maxima seem to
occur near 0.6 (\cite{per90}).  Phase zero is arbitrarily defined to
be at JD $2443366.775$ (\cite{tg82}).  A $\sim 4$ year periodic
modulation has been proposed in the amplitude of the radio maxima
(\cite{gxbr89}), possibly due to precession of a relativistic jet, or
variable accretion from the B star.  The modulation has been fit with
a sinusoidal function with a period of $1476 \pm 7$ d (\cite{per90})
or $1605 \pm 24$ d (\cite{mar93}).  VLBI observations near maxima show
a 2 mas double source implying source size of about $5.5 \times
10^{13}$ cm (3.7 AU) and an equipartition value of 0.7 G for the
magnetic field (\cite{mpef93}).  The source shows evidence for rapid
expansion ($\gtsim 400$ km/s), but not for relativistic bulk motions
such as those seen in GRS 1915+105, GRO J1655$-$40, SS433 and Cyg X-3
(see \incite{mr94} for a review).

In addition, \lsi\ is associated with a weak, variable X-ray source 
which has been observed with {\em Einstein}, {\em ROSAT}, and {\em 
ASCA} (\cite{bcl+81,gm95,typ+96,lhy97}).  The X-ray source (in the one 
well-sampled cycle) is brightest ($\sim 1.5 \times 10^{33}$ erg/s, 
0.07-2.48 keV, assuming a distance of 2.0 kpc and without correcting 
for absorption) during the radio quiet phase, although the 
relationship is poorly determined because of very sparsely sampled 
X-ray data.  The X-ray spectrum is an absorbed power-law with photon 
index $\sim -1.7$ with no evidence for an Iron line (\cite{lhy97}).

\lsi\ is exceptional among X-ray binaries in that it is the most
probable counterpart to the COS-B and EGRET $\gamma$-ray source
2CG\,135+01 (\cite{hsb+77,fbc+94,tfr+96}) with an implied luminosity
of $\sim 8 \times 10^{34}$ erg/s above 100 MeV (for a distance of 2.0
kpc).  The radio position is between the 68\% and 95\% confidence
level contours of the best $\gamma$-ray position determination
(\cite{tbd+95,kab+97}) by EGRET.  Both COMPTEL and OSSE have observed
emission from this region (\cite{vdbb+96,ssc+97}) but have not
been able to exclude the nearby QSO0241+622 as the source.


In 1994 January, we began dual-frequency radio monitoring observations
of \lsi\ to follow the radio flux density and spectral index of the
source during {\em ASCA} X-ray observations (\cite{lhy97,rfw+96b}).  These
observations were continued through 1996 February and supported
{\em Compton Gamma-Ray Observatory} observations (\cite{tfr+96}).
Multi-frequency observations are a good way to probe the physical
conditions of the source.  Because {\mt of the large cycle-to-cycle
variations}, it is imperative to track the state of the radio emission
during high-energy observations.  This paper summarizes the radio
observations (\S\ref{sec-obs}) and discusses implications for models
(\S\ref{sec-disc}) of this unique system.

\section{Observations}
\label{sec-obs}

The observations reported here were made several times per day (with a 
few gaps) from 1994 January 27 to 1996 February 23.  The monitoring 
was performed with the National Radio Astronomy Observatory Green Bank 
Interferometer (GBI) in Green Bank, West Virginia.  The GBI consists 
of two 26 m antennas on a 2.4 km baseline each of which has a pair of 
cooled receivers to simultaneously receive signals at 2.25 and 8.3 
GHz.  The correlators process 35 MHz of bandwidth at each frequency.  
After being shut down due to lack of funding from 1996 April through 
November, the GBI is currently being operated by NRAO and 
NASA primarily for radio monitoring of X-ray sources.  All data taken 
is made available to the public immediately via the WWW ({\tt 
http://www.gb.nrao.edu/gbint/GBINT.html}).  

Ten minute scans on \lsi\ were performed 1--10 times per day within
$\pm 5$ hours of source transit.  Each scan consists of a vector sum
of the correlator amplitude and phase from 30 s integrations.
Measured correlator amplitudes are converted to flux densities by
comparison to standard, regularly observed calibrators.  The resulting
data set consists of $6628$ flux density measurements of the source at
two frequencies over 2.07 yr.  Twenty-five outbursts have nearly
complete coverage and three more have partial coverage (due to
hardware or scheduling problems).  The receivers have a system
temperature of $\sim 40$\,K, and individual flux density measurements
have errors of:
\begin{equation}
	\sigma_{2.25 \rm GHz} = \left[(0.006)^2 + (0.014 S_{2.25 \rm
GHz})^2\right]^{1/2} {\rm Jy}, 
\end{equation}
\begin{equation}
	\sigma_{8.3 \rm GHz} = \left[(0.009)^2 + (0.049 S_{8.3 \rm
GHz})^2\right]^{1/2} {\rm Jy}. 
\end{equation}
Note that the constant term has been significantly lowered since 1989
(\cite{fws+87}) by the installation of new cooled HEMT receivers.  A
fraction of this error estimate is due to a difficult to calibrate
systematic gain variation in the system as a function of hour angle. 
The gain variations are likely due to a combination of changing
aperture efficiency due to deformations of the antenna surface, and
weather-dependent phase coherence problems.  We have attempted to
correct for the gain variations by fitting the hour-angle dependent
effect observed in a constant source with a second order polynomial. 
This was used to correct the observed fluxes for \lsi.  The magnitude
of the effect at $\pm 5^{\rm h}$ is about 7\% and 20\% for 2.25 and 8.3
GHz, respectively.

The daily average flux density of {\lsi} at 2.25 and 8.3 GHz as well
as the calculated spectral index are shown in Figure~\ref{fluxhist}
and \ref{fluxhistb}.  The error bars on this and other figures are
$\pm 1 \sigma$.  There is considerable variation from outburst to
outburst, but inspection of the light curves reveals that the higher
frequency tends to peak earlier and fade more quickly than the lower
frequency.  By cross correlating the light curves at each frequency,
we measure a mean time delay between the 8.3 GHz outburst and the 2.25
GHz outburst of $0.4$ days with a large RMS scatter of 0.5 days.  In
addition, many of the outbursts show two or more peaks during the
decline of the main outburst.

\subsection{Spectral evolution}

\begin{figure}
\caption{Phase averaged fluxes and spectral index of \lsi.  }
\vspace{0.5in}
\psfig{figure={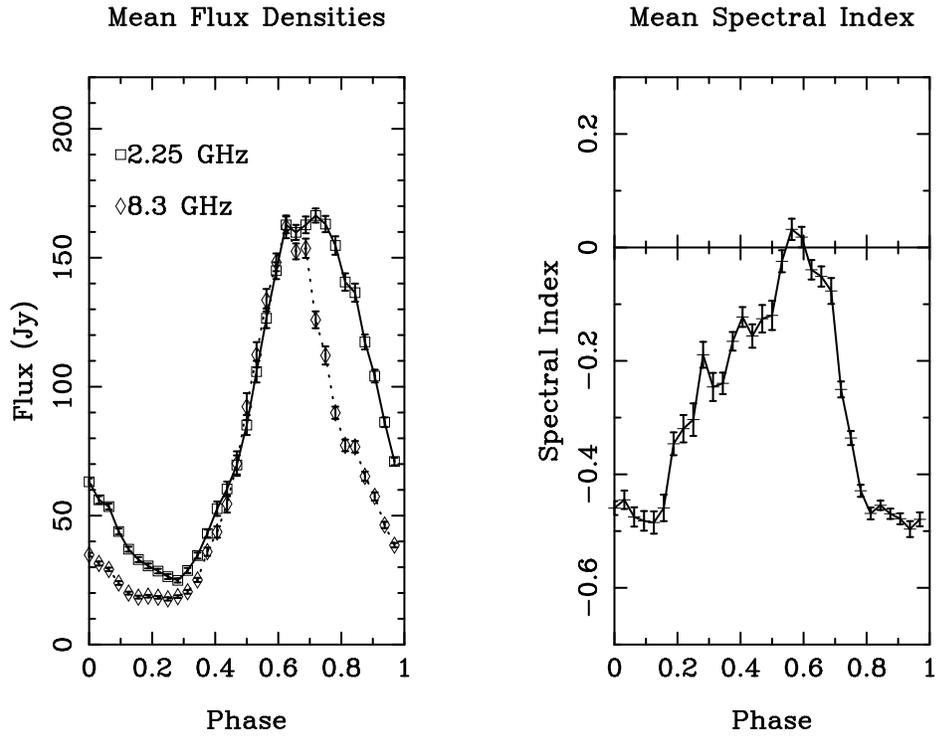},width=5.5in,angle=270}
\label{phaseavg}
\end{figure}

An important feature of this monitoring program is the simultaneous
dual-frequency observations.  This allows accurate measurements of the
spectral index as a function of time (orbital phase) in this highly
variable source.  From two frequency data, we cannot tell anything
about the {\em shape} of the spectrum, but we can infer a spectral
index ($\alpha$) assuming that $S_\nu \propto \nu^\alpha$.  During
typical outbursts we observe the spectral index to rise to a peak with
$\alpha$ between 0 and 0.5 then decay back to the quiescent value of
$\alpha \sim -0.4$.  Our mean light curves with derived two-frequency
spectral index between 2.25 and 8.3 GHz are shown in
Figure~\ref{phaseavg}.  
{\mt Note, however, that the actual radio light curve and 
spectral index behavior may substantially change from cycle to cycle.}
The changing spectral index is a result of the
typical outburst profile decaying more slowly at 2.25 GHz than at 8.3
GHz.  By fitting Gaussian profiles to the outbursts, we find that the
characteristic widths of the 2.25 GHz profiles are 28\% larger than
that of the 8.3 GHz profiles.

The only previous measurements of the spectral index evolution during
an outburst are the statistical results of \incite{tg82} and one
outburst observed by \incite{tg84}.  Both of these results reported
two-frequency spectral indices between 5 and 10 GHz.  Realistic models
for the radio spectrum of this source are not simple power laws
(\cite{mar93}), and free-free absorption will be much more important
at 2.3 GHz than at 5 GHz, so direct comparison of our results to
previous ones is somewhat difficult.  However, in the one outburst
with two-frequency measurements, the spectral index is slightly
negative at the peak and the decay phase of the outburst is consistent
with a constant spectral index.  Also, their statistical results show
a strong peak in spectral index of nearly $\alpha = 2$ at around phase
0.4.  We see no evidence for positive spectral indices except at the
peak of the outbursts.  

There are several energy loss mechanisms which may be important in an
expanding cloud of relativistic electrons including adiabatic
expansion, synchrotron emission, and inverse Compton scattering
(\cite{pac70}).  Each of these affects the shape of the electron
energy distribution function and thus the observed spectral index in
different ways.  Inverse Compton scattering and synchrotron losses
both have $dE/dt \propto E^2$ which result in a spectral index
changing toward more negative values during the decay phase of the
outburst.  Adiabatic expansion implies $dE/dt \propto E$ which
results in a constant spectral index, which is not seen in the decay
phase of the outbursts.  The multiple peaks in the light curves of the
outbursts indicate continuing particle injection.  It is clear that
the production of energetic particles in this source cannot be modeled
as a simple delta function of time.

The observed spectral index evolution may also be (at least partially) 
the result of the source being embedded in a high-optical-depth 
envelope such as the ionized stellar wind of the B star.  In this 
case, the emitted radio emission will be modified by the free-free 
opacity of the wind.  This envelope may be completely opaque near 
periastron and the optical depth may remain greater than unity for 
6--22 days (\cite{mpef93}).  Thus, the observed outburst onset may 
occur well after the initial acceleration event.  Since the wind 
opacity is a strong function of frequency (the optical depth $\tau = 
T_e^{-1.5} \nu^{-2.1} {\rm EM}$, where the emission measure ${\rm EM} 
= \int n_e^{2} ds$; \cite{pac70}), one would expect the higher 
frequency to brighten first if this is an important effect, as is 
observed.

\subsection{Long-term periodicity}

The proposed long-term periodicity has previously been fitted with
rather sparse data: 14 peak flux measurements covering only 2--3
periods of the modulation (\cite{mar93}).  Possible explanations include:
variable beaming from a precessing relativistic jet (\cite{gxbr89}), 
variable wind velocity from the B star resulting in a variable
accretion rate onto the neutron star (\cite{mp95}), or precession of
the B star in the neutron star magnetic field (\cite{ln94}).

\begin{figure}
\caption{Peak outburst flux of \lsi.  The dotted line is the best-fit
model of \protect\incite{per90}.  The dashed line is the more recent
model of \protect\incite{mar93}.  Stars and upward pointing arrows
(lower limits) are the previously published data on which the model
was based.  All data beyond day 9000 are from this work.}
\vspace{0.5in}
\psfig{figure={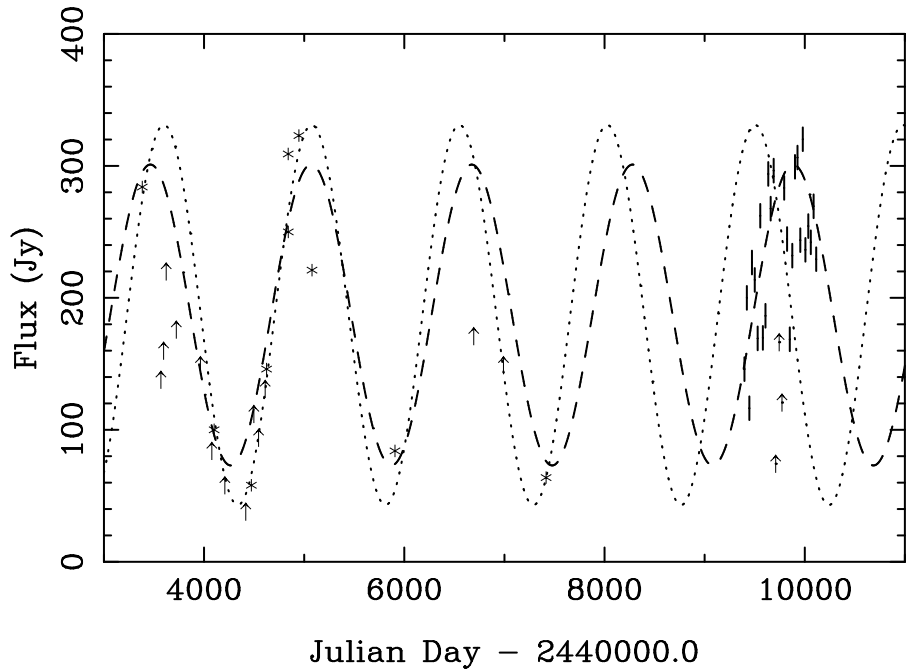},width=5.5in,angle=270}
\label{peakfig}
\end{figure}

\begin{figure}
\caption{Expanded version.  The dotted line is the best-fit model of
\protect\incite{per90}.  The dashed line is the more recent model of
\protect\incite{mar93}.}
\vspace{0.5in}
\centerline{
\psfig{figure={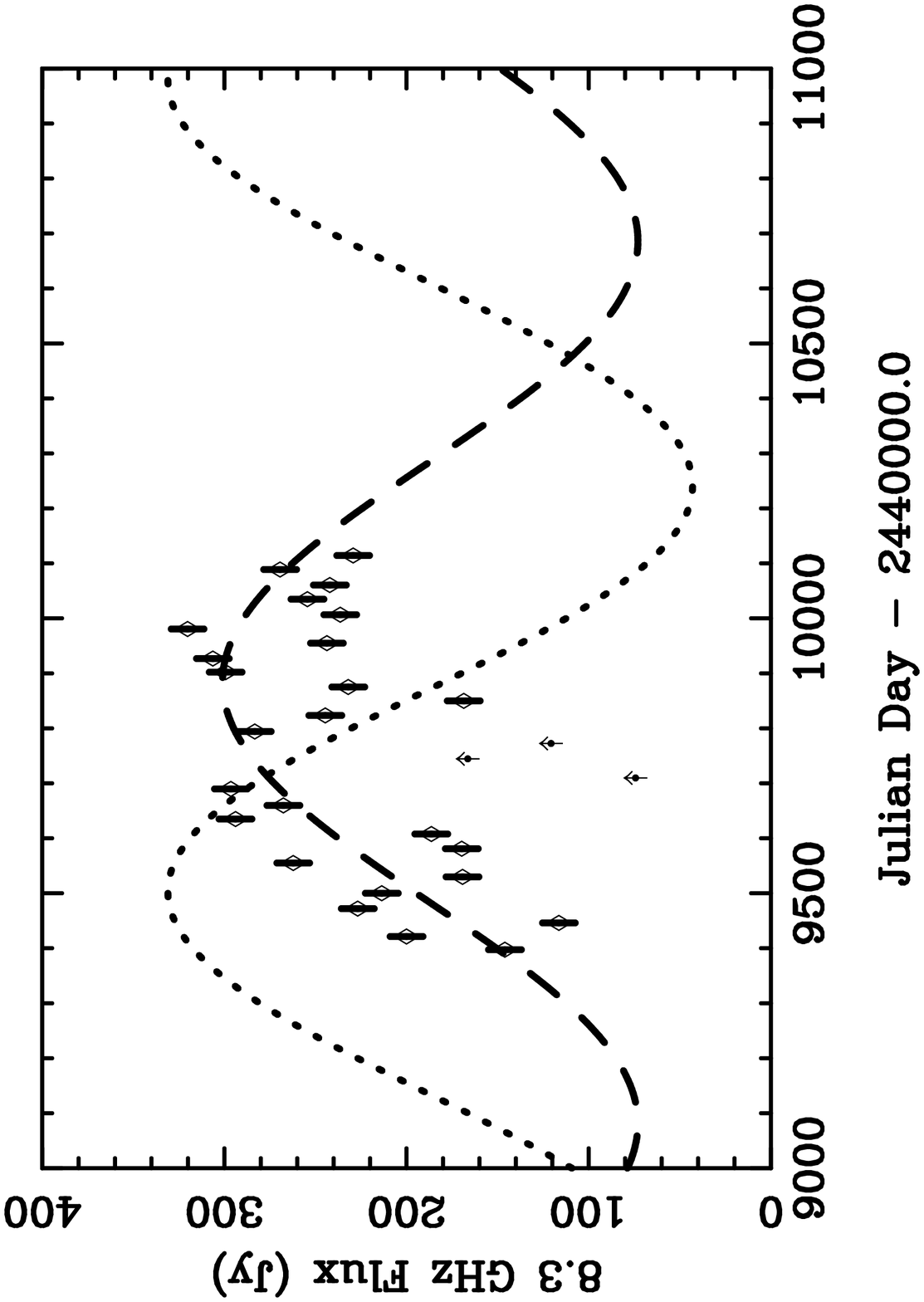},width=5.0in,angle=270}
}
\label{peakfigzoom}
\end{figure}

\begin{figure}
\caption{Same as Figure~\ref{peakfigzoom} except
using the 2.25 GHz data.}
\vspace{0.5in}
\psfig{figure={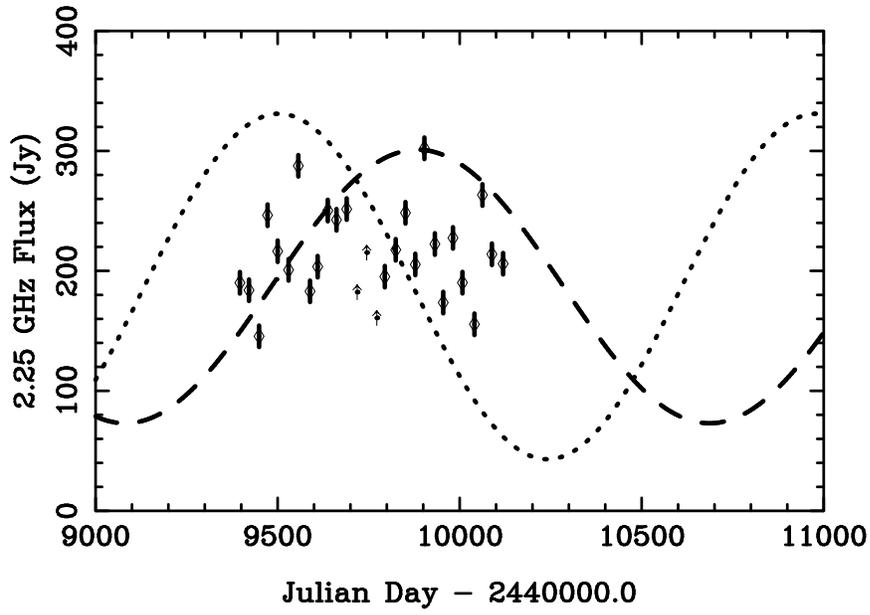},width=5.0in,angle=270}
\label{peakfigzooms}
\end{figure}

The peak fluxes (at 8.3 GHz) of the 18 outbursts observed by the GBI
are shown in Figure~\ref{peakfig} along with the previous observed
outburst maxima (observed at 5, 8, or 10 GHz) used to fit the
sinusoidal model by \incite{per90} and \incite{mar93}.  An expanded
view of only the new data is shown in Figure~\ref{peakfigzoom}.  Both
figures include the previously published models for the modulation.
Our data rule out the four-year modulation model of \incite{per90},
but are marginally consistent with the model of \incite{mar93} in which
\begin{equation}
S_{\rm max} = A \cos \big[\frac{2 \pi}{P} (t-t_0)\big] + B,
\end{equation}
where $A = 114 \pm 8$ mJy, $t_0 = {\rm JD}\: 2443464 \pm 100$, $A =
114 \pm 9$ mJy, and $P = 1605 \pm 24$ d.  However, no evidence for any
secular trend in the peak flux data at 2.25 GHz (shown in
Figure~\ref{peakfigzooms}) is observed.

\subsection{Ephemeris}

\begin{figure}
\caption{Outburst phase determined by cross-correlating
with a template.}
\psfig{figure={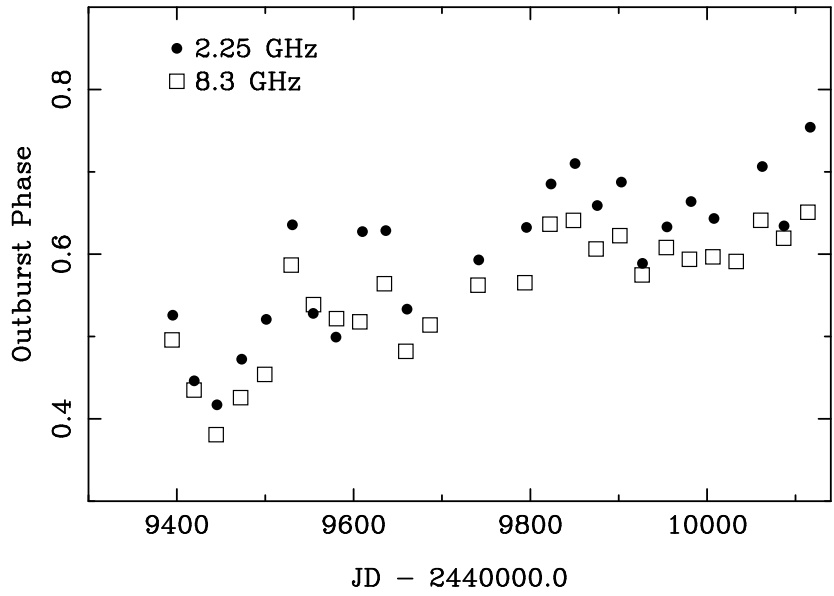},width=5.0in,angle=270}
\label{peakphase}
\end{figure}

Figure~\ref{peakphase} shows the phase of the outburst maxima as a
function of date.  These phases were determined by cross correlating
the observed fluxes with a template.  Due to incomplete sampling of
the fluxes, and occasional high points which may be due to radio
interference, fitting outburst phases by cross correlation is more
robust than simply taking the time of the measurement with the highest
flux.  The offset between the two frequencies may be partially due to
a different template with a different fiducial point being used for
each frequency.

A secular trend in the outburst phases is evident.  To characterize
this effect, we have fit the outburst times of the 8.3 GHz data to a
linear model
\begin{equation}
t_n = t_0 + n \times P,
\label{eq-eph}
\end{equation}
where $t_n$ is the peak time of the $n$th outburst.  Fitting for $t_0$
and P using 26 measured outburst times yields $t_0 = {\rm JD}
2449393.5 \pm 0.4,$ and $P = 26.69 \pm 0.02$ days.  With these
parameters, equation \ref{eq-eph} provides a good ephemeris for recent
outburst maxima.  The RMS difference between the model and the actual
outburst times is 1.1 days.  Since the differences of the measured
outburst times from a linear model could have contributions from many
sources including intrinsic variability, cross correlation errors due
to shape variations from outburst to outburst, and normal measurement
errors, the standard deviations used in the fitting process were
calculated from the RMS residuals after fitting a linear model to the
data.

We note that our best estimate of the period is significantly 
different ($> 9\sigma$) from the best previously published value of 
26.496 $\pm$ 0.008 days (\cite{tg84}).  There are several possible 
explanations for this result.  Either there has been real orbital 
period evolution in the 10--12 years since the previous result was 
published, or the outburst peaks tend to move around in orbital phase 
in response to some other parameter in the system, such as the B star 
mass-loss rate, or a precession period.  This other parameter may also 
be related to the possible modulation in the outburst maxima discussed 
above.  The required orbital period derivative to explain currently 
observed period is $(3.5 \pm 0.03) \times 10^{-5}$, consistent with 
the upper limit of $1.6\times 10^{-4}$ from \incite{tg84}.  The 
timescale implied by $P/\dot{P}$ is $2 \times 10^3$ yr, two orders of 
magnitude less than that observed in the two HMXB systems (Cen X-3 and 
SMC X-1) with measured orbital period derivatives.  Thus the 
interpretation as true orbital period change seems unlikely.  These 
two models predict rather different behavior over the next several 
years.  Real orbital period evolution will result in a continued 
gradual lengthening of the period while a drift in the outburst 
orbital phase predicts that the phase should recover back to 0.6, 
possibly in response to the 4-year modulation.  A complete analysis of 
all published data over the 20 years since the source was discovered 
is in progress (M. Peracaula, personal communication) which may 
resolve this question.

\section{Discussion}
\label{sec-disc}


Most models of this source share the feature that the radio emission
is due to synchrotron emission from relativistic electrons moving in a
magnetic field.  However, there is no consensus as to the means of
production of these particles, or their relation to the high-energy
emission.  Several mechanisms for creating the required relativistic
electron population which have been proposed for {\lsi} are summarized
below to demonstrate the wide variety of possible mechanisms.

\incite{mt81} {\mt (see also \cite{t95})} proposed that the
relativistic electrons are from a young pulsar wind, and that the
X-ray and $\gamma$-ray emission comes from Compton scattering of
optical photons from the primary B star at a shock boundary between
the pulsar wind and the stellar wind.  \incite{tg84} proposed that
super-Eddington accretion onto the neutron star near periastron
results in luminosity-driven shocks which may account for the particle
production.  \nocite{pmes91} Paredes {\etal} (1991) proposed an
adiabatically expanding synchrotron source with prolonged injection of
relativistic particles.  \incite{ln94} proposed that relativistic
electrons from a pulsar wind are captured by the magnetosphere of the
B star and cooled by synchrotron losses.  Recent work (\eg
\cite{csmc95}) has focused on the accretion regimes of rapidly
rotating magnetized neutron stars in eccentric binaries.  Similarly,
\incite{zam95} proposed that the outbursts occur at the transition of
a magnetized neutron star from a propeller phase to an ejector phase
as the mass transfer rate changes in an eccentric orbit.

It is likely that {\lsi} is a close relative of the two recently
discovered radio pulsars in B-star binaries (PSR\,J0045$-$7319 and
PSR\,B1259$-$63) and the 69-ms X-ray pulsar A\,0535$-$668.  All three
systems contain a rapidly rotating, highly magnetized neutron star
orbiting a massive main-sequence companion.  The primary differences
between these systems are the mass-loss rate of the OB star, the
geometry of the orbit, and the energy flux of the pulsar wind.  These
three factors determine which accretion regime the system is in.  For
J0045$-$7319, $\dot M < 10^{-10} \Msun\,{\rm yr}^{-1}$ (\cite{ktm95})
and for B1259$-$63, $100 < (v / 10 {\rm\, km\,s}^{-1}) (\dot M /
10^{-8} \Msun\,{\rm yr}^{-1}) < 10^4$ (\cite{ta97}).  Thus, {\mt these
two systems } have fairly weak stellar winds, active pulsars producing
strong outflows, and B1259$-$63 does not get very close to the
companion at periastron, allowing both to remain in the radio pulsar
regime since matter is unable to accrete onto the {\mt pulsar}
magnetosphere.  On the other hand, A\,0535$-$668, {\mt may have } a
weaker pulsar wind, and might encounter the thick equatorial disk of
the Be star near periastron.  This allows accretion at about the
Eddington rate to occur and produces a powerful X-ray pulsar.  For
\lsi, {\mt a reasonably good estimate for the mass loss rate from the
companion is} $\dot M \sim 10^{-7} \Msun\,{\rm yr}^{-1}$
(\cite{wtv+88}).  \nocite{csmc95} Campana {\etal} (1995) suggest that
while in {\lsi} the stellar wind ram pressure may be sufficient to
overcome the pulsar wind pressure and accrete onto the magnetosphere,
the angular velocity might be super-Keplerian resulting in the matter
being ejected by the propeller mechanism.  {\mt Alternately, a
relatively young pulsar may power the high energy emission from \lsi\
(\cite{mt81,t95}) by a shock mechanism which can be shown to be in
agreement with the intensity and spectral ranges observed for \lsi\ in
the X-ray range and for 2CG\,135+01 in the gamma-ray range
(\cite{tfr+96}).  }

We note that the long-timescale monitoring at 2.25 GHz adds 
significant new information compared to previous observations of \lsi\ 
that were made a much higher frequency.  This can be immediately 
realized by considering the relevance of Figure~\ref{phaseavg} giving 
the mean spectral index.  The flattening of the spectrum at phases 
coincident with the peak radio fluxes is evident.  Free-free 
absorption along the line-of-sight or synchrotron self-absorption may 
be responsible for the spectral flattening near phase 0.6.  If we 
assume a critical absorption frequency at, say, 5 GHz, and a gas 
temperature of $10^4$~K, we obtain a free-free optical depth larger 
than unity for an emission measure $EM \geq 2 \cdot 10^{25}$ (cgs 
units).

We can also estimate the source size $L$ if the spectral flattening is 
caused by synchrotron self-absorption.  For an estimated radio 
luminosity of $\sim 2 \times 10^{30} \rm \, erg \, s^{-1}$, a local 
magnetic field of $\sim 1$~G, we deduce $L \sim 7 \times 10^{12}$~cm.  
We notice that the estimated value of $L$ is not dissimilar from the 
source size adopted in the adiabatic/synchrotron cooling model by 
\cite{pmes91}.  The integration time of 10 minutes prevents us from
placing a strong limit on the source size from rapid variability, but
we do find that the source varies by more than a factor of two within
15 minutes.  This is fully consistent with the timescale $L/c \sim 4$~min.

\section{Summary}

After 20 months of continuous radio observations of \gt\ we find only
weak evidence for long-term {\mt modulation of peak flare amplitudes}
on a four year time scale.  We observe spectral index variability
during the outbursts which implies that either adiabatic expansion is
not the dominant energy loss mechanism in the expanding plasmon or
that the optical depth due to free-free opacity is changing during the
outburst.  We also observe an apparent lengthening of the period
between outbursts which is fit with a period of $26.69 \pm 0.02$ days.
This may be due to an orbital period derivative of $(3.5 \pm 0.03)
\times 10^{-5}$ or due to the outburt phases being modulated in
response to some other parameter of the system.

A single self-consistent model for \lsi\ remains elusive.  Critical
questions include: Are the X-rays and $\gamma$-rays produced by the
same population of relativistic electrons? Is the compact object a
young pulsar? What is the orbital eccentricity? 

A critical diagnostic for distinguishing between the various models is 
the time dependence and correlation between the radio, X-ray and 
$\gamma$-ray emission.  Multiwavelength observations with {\em GRO}, 
{\em XTE}, and {\em SAX} will be important.  A collection of resources 
to facilitate multifrequency observations of this source is available 
on the WWW at {\tt 
http://www.srl.caltech.edu/personnel/paulr/lsi.html.} Radio 
observations at more than two frequencies will also be useful to 
separate the effects of free-free absorption, synchrotron and inverse 
Compton losses in the evolution of the source spectrum.  Finally, a 
more accurate determination of the Keplerian orbital parameters is 
badly needed.

\acknowledgments

During this work, the Green Bank Interferometer was operated by the 
National Radio Astronomy Observatory for the Naval Research Laboratory 
and was supported by the Office of Naval Research.  Partial support for 
the observations and analysis of this source were supported by a 
National Aeronautics and Space Administration interagency fund 
transfer.  Basic research in precision pulsar astrophysics at the 
Naval Research Laboratory is supported by the Office of Naval 
Research.  A portion of this work was performed while one of the 
authors (PSR) held a National Research Council-NRL Research 
Associateship.  This research made use of the Simbad database, 
operated at CDS, Strasbourg, France.

\clearpage


\bibliographystyle{apj}

\clearpage
\begin{figure}
\caption{Flux history of \lsi\ at 2.25 and 8.3 GHz from 27 January
1994 to 8 March 1995.  The arrows mark phase 0.6 of the published
radio ephemeris (\protect\cite{tg84}).  The triangles mark the
outburst phase as determined by cross-correlating with a template (see
Figure~\ref{peakphase}).  The triangles are missing where there was
insufficient data to get a good peak.}
\vspace{0.5in}
\centerline{
\psfig{figure={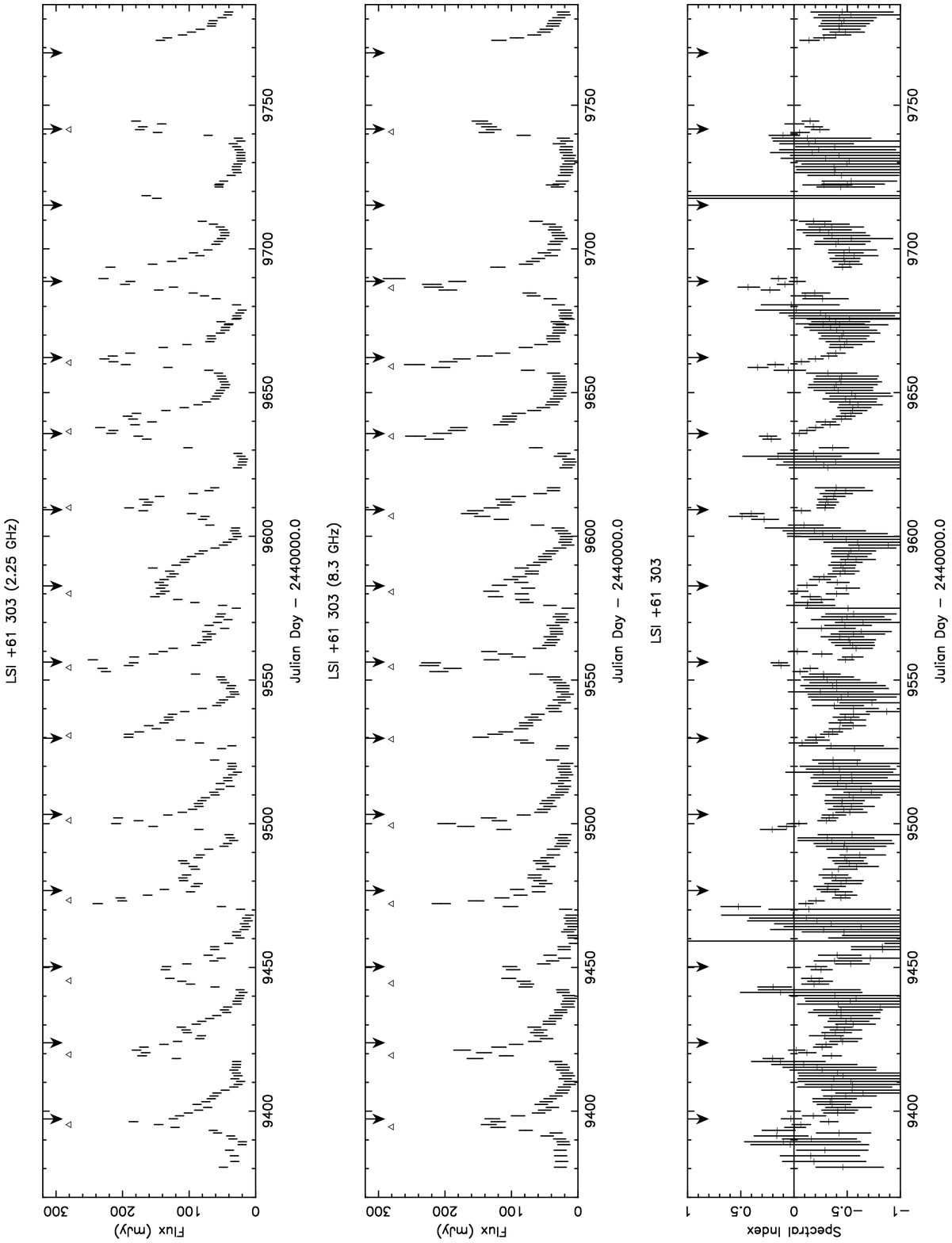},width=5.0in}
}
\label{fluxhist}
\end{figure}

\begin{figure}
\caption{Continuation of Figure\ref{fluxhist} from 8 March 1995 to 26
February 1996.}
\vspace{0.5in}
\psfig{figure={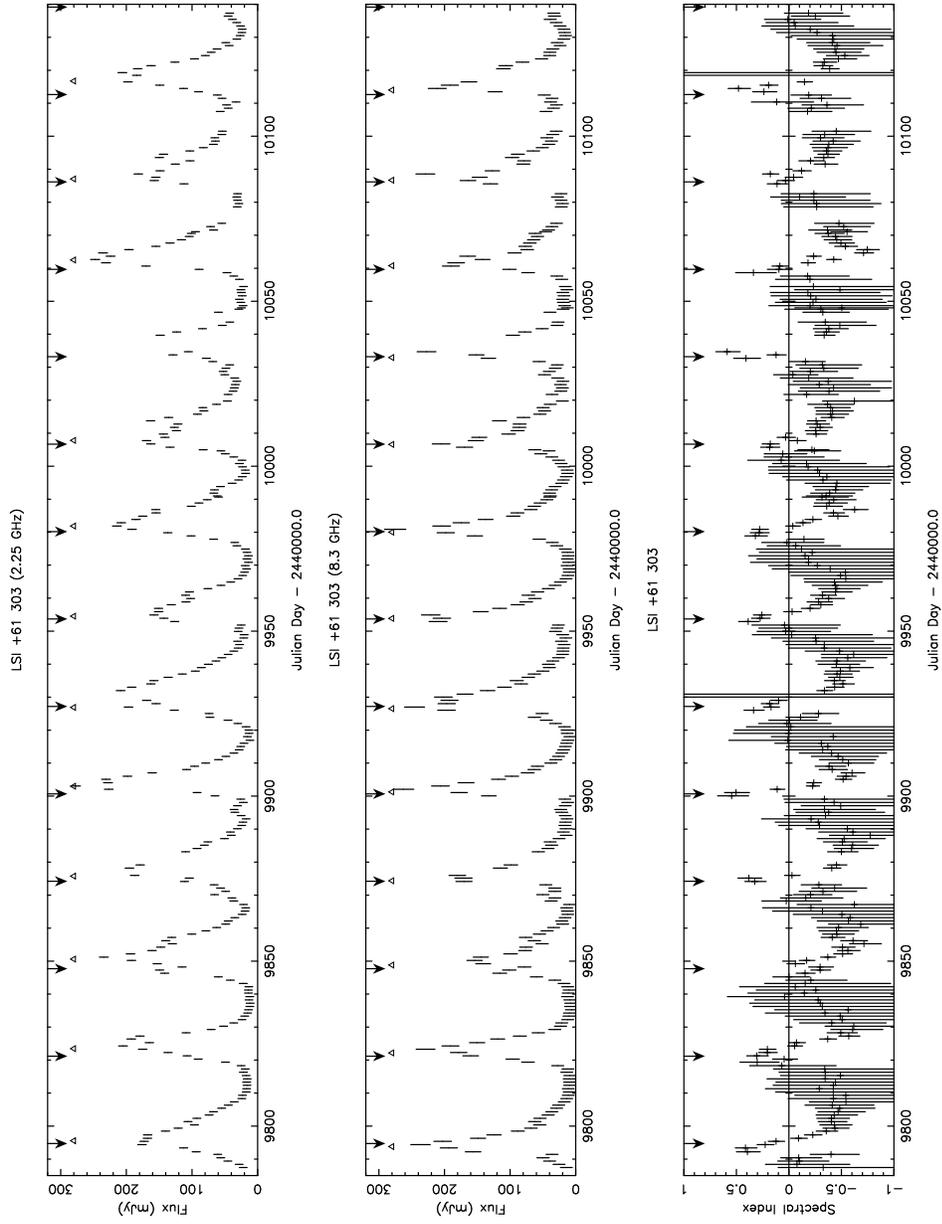},width=5.0in}
\label{fluxhistb}
\end{figure}

\end{document}